\documentclass[manuscript,screen,nonacm]{acmart}
\acmConference[]{} 

\newcommand{\workshopname}{GenAICHI: CHI 2024 Workshop on Generative AI and HCI}
\newcommand{\licensedetails}{Licensed under a Creative Commons Attribution 4.0 International License (CC BY 4.0). Copyright remains with the author(s).}
\newcommand\extrafootertext[1]{
    \bgroup
    \renewcommand\thefootnote{\fnsymbol{footnote}}%
    \renewcommand\thempfootnote{\fnsymbol{mpfootnote}}%
    \footnotetext[0]{#1}%
    \egroup
}

\AtBeginDocument{ 
    \fancypagestyle{firstpagestyle}{
        \fancyhf{}
        \fancyfoot[L]{\sffamily\footnotesize \workshopname}%
        \fancyfoot[C]{\sffamily\footnotesize \thepage}
    }
    \fancyhf{}
    \fancyhead[L]{\sffamily\footnotesize\shorttitle}
    \fancyhead[R]{\sffamily\footnotesize\shortauthors}
    \fancyfoot[L]{\sffamily\footnotesize\workshopname}%
    \fancyfoot[C]{\sffamily\footnotesize\thepage}
    \extrafootertext{\licensedetails}
}

\AtBeginDocument{%
  \providecommand\BibTeX{{%
    \normalfont B\kern-0.5em{\scshape i\kern-0.25em b}\kern-0.8em\TeX}}}

\setcopyright{acmlicensed}
\copyrightyear{2024}
\acmYear{2024}
\acmDOI{10.17605/OSF.IO/FVTMQ}

\acmConference[GenAI Workshop '24]{GenAI Workshop '24}{May 11--16,
  2024}{Honolulu, Hawaii}
\acmBooktitle{GenAI Workshop '24 May 11--16, 2024 Honolulu, Hawaii} 
\acmISBN{10.17605/OSF.IO/FVTMQ}

\usepackage{comment}
\usepackage{float}

\usepackage{xcolor,colortbl}
\definecolor{Hildaspink}{HTML}{FF99BE}
\definecolor{red}{HTML}{FF0000}

\usepackage{enumerate}  
\usepackage[shortlabels]{enumitem}

\usepackage{geometry}
\usepackage[utf8]{inputenc}

\begin{document}

\title[Augmenting the Author]{Augmenting the Author: Exploring the Potential of AI Collaboration in Academic Writing}

\author{Joseph Tu}
    \authornote{Authors are also affiliated with the Department of Systems Design Engineering, University of Waterloo}
    \email{joseph.tu@uwaterloo.ca}
    \orcid{0000-0002-7703-6234}
    \affiliation{
    \institution{Stratford School of Interaction Design and Business, University of Waterloo}
    \streetaddress{200 University Ave W}
    \country{Canada}
}

\author{Hilda Hadan}
\authornotemark[1]
\email{hhadan@uwaterloo.ca}
\orcid{0000-0002-5911-1405}
\affiliation{
    \institution{Stratford School of Interaction Design and Business, University of Waterloo}
    \city{Waterloo}
    \country{Canada}
}

\author{Derrick M. Wang}
\authornotemark[1]
\email{dwmaru@uwaterloo.ca}
\orcid{0000-0003-3564-2532}
\affiliation{
    \institution{Stratford School of Interaction Design and Business, University of Waterloo}
    \city{Waterloo}
    \country{Canada}
}

\author{Sabrina A. Sgandurra}
\email{sasgandu@uwaterloo.ca}
\orcid{0000-0003-3187-263X}
\affiliation{
    \institution{Stratford School of Interaction Design and Business, University of Waterloo}
    \city{Waterloo}
    \country{Canada}
}

\author{Reza Hadi Mogavi}
\email{rhadimog@uwaterloo.ca}
\orcid{0000-0002-4690-2769}
\affiliation{
  \institution{Stratford School of Interaction Design and Business, University of Waterloo}
  \city{Waterloo}
  \state{Ontario}
  \country{Canada}
  }  

\author{Lennart E. Nacke}
\authornotemark[1]
\email{lennart.nacke@acm.org}
\orcid{https://orcid.org/0000-0003-4290-8829}
\affiliation{
    \institution{Stratford School of Interaction Design and Business, University of Waterloo}
    \city{Waterloo}
    \country{Canada}
}

\renewcommand{\shortauthors}{}

\begin{abstract}
This workshop paper presents a critical examination of the integration of Generative AI (Gen AI) into the academic writing process, focusing on the use of AI as a collaborative tool. It contrasts the performance and interaction of two AI models, Gemini and ChatGPT, through a collaborative inquiry approach where researchers engage in facilitated sessions to design prompts that elicit specific AI responses for crafting research outlines. This case study highlights the importance of prompt design, output analysis, and recognizing the AI's limitations to ensure responsible and effective AI integration in scholarly work. Preliminary findings suggest that prompt variation significantly affects output quality and reveals distinct capabilities and constraints of each model. The paper contributes to the field of Human-Computer Interaction by exploring effective prompt strategies and providing a comparative analysis of Gen AI models, ultimately aiming to enhance AI-assisted academic writing and prompt a deeper dialogue within the HCI community.
\end{abstract}



\received{26 February 2024}
\received[accepted]{13 March 2024}

\maketitle

\section{Introduction}\label{intro}
In the realm of scholarly investigation, traversing the preliminary stages of drafting a research manuscript can pose challenges~\cite{sedlmair2012design}. Some of the most common struggles when writing research papers are the tight schedule, lack of inspiration (for writing), and the complex process of organizing ideas and creating an outline~\cite{juzwik2006writing,almarwani2020academic}. This issue is particularly salient for early career academics and individuals whose first language is not English~\cite{zhao2023leveraging,fitria2021grammarly}. We ask ourselves, could Artificial Intelligence (AI) help?

The purpose of our workshop paper is to explore the potential of using AI to address challenges faced by the academic research community. Through an exploratory research approach, we aim to gather intellectual feedback and initiate a dialogue within the HCI community. Instead of presenting a traditional or formal research paper, our goal is to open a discussion on this topic. We will be sharing our combined reflections on the use of AI, focusing on two popular Generative AI models - ChatGPT\footnote{ChatGPT.~\url{https://chat.openai.com/share/bf65d447-70c8-4117-87ef-f258265e3e43}} and Gemini\footnote{Gemini.~\url{https://gemini.google.com/share/259d4d8f69f8}}.

However, the application of AI as a writing assistant is often impeded by a dearth of transparency in researchers' access and utilization ~\cite{rai2020explainable, balasubramaniam2023transparency}. This lack of transparency raises concerns about the reliability and credibility of AI-generated text~\cite{ali2020artificial, baidoo2023education}. Hence, comparative analysis between ChatGPT and Gemini as writing assistants for research draft outlines becomes crucial in understanding their capabilities and potential limitations.

At its core, our research emphasizes the importance of continual reflection and validation throughout the entire process of interacting with Gen AI. This includes thoughtful, prompt design, comprehensive output analysis, and an awareness of potential limitations. We contend that prioritizing this process-oriented approach is crucial for responsibly integrating Gen AI into academic research, leading to more accurate and reliable outcomes for researchers.

For this workshop paper, our discussion will revolve around two key aspects:
\begin{itemize}
  \item\textbf{Exploring Effective Prompt Strategies:} In this section, our attention is directed towards the realm of prompts that researchers themselves generate (the authors), as well as the pivotal inquiries that guide the outputs of Gen AI models, including ChatGPT and Gemini. Drawing upon the valuable insights acquired from our weekly deliberations, we aim to distinguish and articulate efficient strategies for crafting prompts.

  \item \textbf{A Comparative Perspective on Results:} We compare two Gen AI models by giving them the same prompts and asking them to generate outlines for a specific research topic. Our collaborative inquiry approach will reveal the nuanced differences in their responses.
\end{itemize}

As there is growing research in prompt engineering, we aim to expand upon existing contributions, including recent works from \cite{ekin2023prompt,rodriguez2023prompts,wang2023unleashing}, by undertaking a thorough examination of the merits and drawbacks of different models in their handling of various prompt methods. Furthermore, we present a detailed account of the collaborative process with AI, providing the researcher's perspective and promoting transparency and accountability in assessing the efficacy and limitations of ChatGPT and Gemini as writing aids. We focus not on formal and statistical comparisons, as we acknowledge potential subjective biases and confounding factors. Instead, we aim to offer a firsthand experience that can serve as a starting point for meaningful discussions within the HCI community, potentially leading to further and more comprehensive investigations.

This position paper aims to make two significant contributions to the field of HCI. Firstly, it seeks to delineate effective prompt engineering strategies that enhance the collaborative dynamic between human researchers and AI, thereby optimizing the integration of AI in the research outline development process. Secondly, the paper aims to provide a nuanced comparative analysis of the capabilities and limitations of two prevalent generative AI models, ChatGPT and Gemini, grounded in real-world application scenarios that could serve as a foundational framework for future research and development in AI-assisted academic writing.

\section{Research Approach}
We used a collaborative inquiry approach~\cite{bray2000collaborative}, focusing on facilitating structured and inclusive discussions and deliberations among ourselves as researchers; and culminating in us collectively reaching thoughtful judgments aligned with a specific goal or purpose, such as formulating recommendations or crafting a research outline. This case study aimed to evaluate the strengths and limitations of large language models (LLM), particularly Gemini and ChatGPT. Researchers (4---6) participated in weekly sessions, utilizing Otter.Ai for transcription. Discussions focused on the strengths and limitations of the models, guided by carefully crafted prompts. The strategic use of AI models in research outline development was evaluated based on predefined criteria. AI-generated prompts further facilitated critical thinking and reflection (condensed for a workshop see \autoref{tab:collaborative_inquiry}).

\begin{table*}
\centering

\caption{An overview of our collaborative inquiry approach}
\label{tab:collaborative_inquiry}
\resizebox{\textwidth}{!}{
\begin{tabular}{|p{10em}|p{50em}|} 
\hline
\rowcolor[rgb]{0.749,0.749,0.749} \textbf{Approach} & \textbf{Description} \\ 
\hline
A. Weekly Collaborative Inquiry Sessions & \begin{tabular}[t]{@{}p{50em}@{}}\textbf{Format and Frequency}: Engaging in regular weekly collaborative inquiry sessions, initiated on January 19th (still on-going as of this submission), where 4-6 researchers convene for one-hour discussions. These sessions provide a structured platform for collective exploration and dialogue.\\\textbf{\textbf{Transcription}}: We used Otter.Ai as a recording tool during these sessions, fostering collaborative transcription and documentation. This software not only ensures accuracy in capturing discussions but also facilitates easy reference and analysis of key points for us.\end{tabular} \\ 
\hline
B. Collaborative Inquiry Prompts and Guiding Questions & \begin{tabular}[t]{@{}p{50em}@{}}\textbf{Designed Prompts}: The research design incorporates carefully crafted prompts. These prompts are strategically formulated to extract nuanced insights from the collective group, emphasizing the strengths and limitations of each AI model.\\\textbf{\textbf{Guidance in Collaborative Deliberations}}: These prompts play a pivotal role in steering collaborative discussions. By focusing on specific aspects, the prompts encourage a detailed examination of the AI models, fostering a deeper understanding of their capabilities in the context of outlines for research papers.\end{tabular} \\ 
\hline
C. Collaborative Research Outline Development & \begin{tabular}[t]{@{}p{50em}@{}}\textbf{Strategic Utilization of AI Models}: At the core of the collaborative inquiry process is the strategic use of both Gemini and ChatGPT in generating research outlines. We actively explore how these AI models collectively structure and organize their ideas coherently, aiming to streamline the initial stages of the research process.\\\textbf{\textbf{Criteria for Collaborative Evaluation}}: The effectiveness of each AI model in research outline development is collaboratively evaluated based on predefined criteria. These criteria include the coherence of the generated outlines, alignment with research objectives, and overall recommendations and critical questions planned for the research phase.\end{tabular} \\ 
\hline
D. Collaborative Critical Thinking Facilitation & \begin{tabular}[t]{@{}p{50em}@{}}\textbf{Collaborative Process of AI-Generated Prompts}: To enhance critical thinking among the research team. We asked the tools to provide us with critical thinking questions and suggestions to enhance the writing process. These prompts are specifically designed to stimulate collective reflection and provoke thoughtful analysis of the research topics under consideration.\\\textbf{\textbf{Objectives of Collaborative Analysis}}: The deliberate facilitation of critical thinking through AI prompts aims to assess the intellectual stimulation provided by Gemini and ChatGPT. It also guides researchers collaboratively in refining their research objectives through a more thoughtful and analytical approach.\end{tabular} \\ 
\hline
E. Collaborative Perspective Elicitation & \textbf{Individual Insights and Their Reflections}: The research methodology includes individual insights as a means to gain diverse perspectives on Gemini and ChatGPT. These reflections serve as a targeted approach to understanding how each researcher collectively perceives the strengths and limitations of the respective AI models. \\
\hline
\end{tabular}
}

\end{table*}

\section{Preliminary Analysis of ChatGPT and Gemini: Our Collective Reflection.}

Through a series of comparisons and observations, we identified key differences in their approach, information handling, and potential applications within the research process. Exploring the capabilities of ChatGPT and Gemini in various tasks prompts a critical examination of their strengths and limitations within the context of research outline development. While ChatGPT consistently fulfills tasks based on explicit instructions, concerns arise due to its tendency to fabricate responses, raising questions about the reliability of generated content~\cite{GPT42023}. We often noted that the model's word-by-word rephrasing approach, coupled with a reluctance to restructure sentences autonomously, emphasizes its dependence on precise directives. Both models occasionally struggle with LaTeX formatting, which introduces complexities that may impact their usability in crafting research outlines (at the time of this submission).  
On the other hand, Gemini proves to be a dynamic counterpart, not only completing tasks but also providing valuable suggestions and rationales. Its ability to guide users through reference searches enhances the reliability of information retrieval. Gemini's capacity to offer multiple word choice suggestions, each accompanied by detailed explanations, adds sophistication to its linguistic capabilities. In tasks requiring paragraph improvement based on references, both models demonstrate a dependence on explicit prompts as outlined by~\citet{lo2023clear}, highlighting the importance of precise instruction in research outline development.

\begin{table*}[ht!]
  \centering
    \caption{A preliminary summary of our collective reflection findings for ChatGPT and Gemini}
  \label{tab:summary}
   \resizebox{\textwidth}{!}{
  \begin{tabular}{|p{10em}|p{25em}|p{25em}|}
    \hline
    \rowcolor{lightgray}
    \textbf{Category} & \textbf{ChatGPT Preliminary Findings} & \textbf{Gemini Preliminary Findings} \\
    \hline
    \textbf{Strengths} & Consistent adherence to explicit instructions, precision, and methodical approach. & Dynamic, provides additional suggestions and rationales. Proficient in guiding through reference searches. Offers multiple word choice suggestions with detailed explanations. \\
    \hline
    \textbf{Limitations} & Tendency to fabricate responses, reliance on word-by-word rephrasing, occasional struggle with LaTeX formatting. & Reliance on explicit prompts in tasks requiring paragraph improvement based on a reference. Limited customization options compared to ChatGPT. \\
    \hline
    \textbf{Articulating AI Potential} & More straightforward outlines. & Excels in comprehensive summaries with self-promotion. \\
    \hline
    \textbf{Handling Mistakes} & May fabricate information, raising concerns about reliability. & Prioritizes transparency, explains confusing sources, and guides on independent information retrieval. \\
    \hline
    \textbf{Human-AI Collaboration} & Ongoing exploration and adaptation required for maximizing benefits and mitigating limitations in the evolving dynamics. & Ongoing exploration and adaptation are required to maximize benefits and mitigate evolving dynamics' limitations. \\
    \hline
    \textbf{Common Sense Integration} & Deficiency in common sense could be addressed through customization using its panel. & Undetermined \\
    \hline
    \textbf{Proactivity} & Undetermined & Demonstrates initiative and understanding of broader context, delivering optimal outcomes without waiting for specific directives. \\
    \hline
    \textbf{AI's Role in Education} & \multicolumn{2}{p{50em}|}{Reflects on AI's potential evolution into a personalized learning tool that adapts to individual learning styles.} \\
    \hline
  \end{tabular}
  }
\end{table*}

In terms of articulating the potential of generative AI for research outlines, Gemini distinguishes itself with comprehensive summaries that include self-promotion. In contrast, ChatGPT tends to provide more straightforward summaries without articulating into additional insights. We also identified ChatGPT's potential limitations in discerning the best method or meaning, and found its potential deficiency in common sense. This suggest that enhancing ChatGPT's performance could be achieved through greater customization using its customization panel, potentially infusing a form of common sense into its responses, particularly when generating content for research outlines. 

In addition, Gemini impresses with its proactive and initiative-taking nature. It demonstrates a keen understanding of the broader context and consistently suggests new phrasings or alternative options without explicit requests, showcasing its potential to aid research outline development without waiting for specific directives. A notable distinction lies in how these models handle mistakes. While ChatGPT at times resorts to fabricating information, raising concerns about reliability and inadvertent alterations of intended meaning, Gemini prioritizes being helpful over maintaining an appearance of infallibility. Gemini transparently addresses limitations, enhancing its credibility and transparency compared to ChatGPT, particularly in the context of research outline development.

\textbf{Researcher's Antidote on Certain Concerns.} 
\textit{Participant 1 (P1)} quotes: ``\textit{This worries me as a reviewer, am I a human validating the author's work or AI's work?}'' One notable concern is the potential shift in the role of reviewers, who may unintentionally become the primary validators of AI-generated content rather than the authors themselves. The human touch, creativity, and critical thinking are essential components of writing. If AI takes a prominent role in generating content, there is a risk of diminishing these human qualities. We suggest authors utilizing AI in their writing should prioritize transparency by clearly disclosing the involvement of AI tools in the creation process. 

\textit{P6} quotes: ``\textit{After using these tools for a while, I have a bias towards a certain writing style. I can tell that AI writes this or that, but what happens if this is just their writing style? Personally, English as a second language, I was taught to write in a formal construct.}'' 
As we explore the potential of AI writing tools, we acknowledge concerns surrounding the negative perception of their generated styles, particularly within academia. We understand the apprehension that might stem from unfamiliarity with this format, potentially leading to bias against its lack of established conventions. Additionally, we recognize the fears surrounding the potential for both quality concerns and ethical misuse of these tools. These concerns could translate into various challenges. We might encounter resistance towards adopting AI writing tools, creating an uneven playing field for those with access. This further highlights the need for clear guidelines to ensure responsible and ethical usage within academic settings \cite{HadiMogavi2024}. While addressing these concerns is crucial for successful integration, we believe that continuous development and increasing familiarity with AI writing might eventually alleviate the initial negative connotations.

\begin{acks}
This workshop was supported by the SSHRC INSIGHT Grant (grant number:435-2022-0476), NSERC Discovery Grant (grant number: RGPIN-2023-03705) and CFI John R. Evans Leaders Fund or CFI JELF (grant number: 41844). Special thanks to the members of the Games Institute at the University of Waterloo for all their support. 
\end{acks}

\bibliographystyle{ACM-Reference-Format}
\bibliography{reference}

\end{document}